# MOND's Problem in Local Group

Yan-Chi Shi [1]

## ABSTRACT

I use the distances and motions of Local Group galaxies to test Modified Newtonian Dynamics (MOND). The old Local Group timing argument of Kahn & Woltjer, which assumed Newtonian gravity and a simple radial orbit, indicated that the total mass of M31 and the Milky Way far exceeds their known baryonic mass. Here I apply MOND to study the same problem. With the same radial orbit assumption, I find that the total mass of M31 and the Milky Way predicted by MOND is less than the known baryonic masses of these two galaxies. I find a similar result holds for all the dwarf galaxies in the Local Group, if they are assumed to move radially with respect to the center of mass of M31 and the Milky Way. If the known baryonic mass of M31 and Milky Way is used, MOND requires *all* galaxies in the Local Group to have non-radial motions with respect to the center of mass of M31and the Milky Way in order to be consistent with their observed motion and distance. The non-radial orbit requirement is inconsistent with the requirement of high radial anisotropy in order to reconcile MOND with the velocities of satellite galaxies around host galaxies from the Sloan Digital Sky Survey.

*Subject headings:* dark matter – Local Group – galaxies: interactions – galaxies: kinematics and dynamics

## 1. INTRODUCTION

There is strong evidence for the existence of dark matter in galaxy clusters and on galaxy scales (Zwicky 1933; Rubin & Ford 1972). Dark matter also plays a key role in our current idea for the growth of large-scale structure in the universe (e.g. Springel et al. 2005). At the largest scale, the flatness of the observed universe and the cosmic microwave background data requires 23 per cent of the mass-energy density in the universe to be dark matter (Komatsu et al. 2008). However, attempts to detect dark matter directly have failed, despite decades of effort.

Newton's law of gravitation and the theory of general relativity are well tested on small scales but departures from these laws on larger scales could eliminate the need for dark matter. The most successful alternative to dark matter is Modified Newtonian Dynamics (MOND), which Milgrom (1983) based on two well-established properties of spiral galaxies: flat rotation curves and the Tully-Fisher relation (Sanders & McGaugh 2002). His proposal is purely phenomenological, but has gained deserved notice from its ability to predict the detailed shapes of disk galaxy rotation curves (Sanders 2008).

[1] E-mail: yanchishi@yahoo.com

MOND is extremely successful in eliminating mass discrepancies in spiral galaxies, since it requires only the baryonic mass to account for their flat rotation curves (Sanders & McGaugh 2002). Some tests of MOND (Gerhard & Spergel 1992; Pointecouteau & Silk 2005; Funkhouser 2005; Slosar et al. 2005; Klypin & Prada 2009) weakly favor dark matter, but the most convincing evidence is the recent gravitational lensing study of 1E 0657-558 also known as the Bullet Cluster (Clowe et al. 2006). This study of two colliding galaxy clusters finds a mismatch between the projected positions of the peak baryonic density and of the total mass estimated by weak gravitational lensing. Even this case, some unseen mass, such as massive neutrinos (Angus et al. 2007) or dark baryons (Milgrom 2008), could possibly save MOND.

Here, I test MOND by extending the old timing argument for the Local Group originally proposed by Kahn & Woltjer (1959). I find that either MOND predicts the total mass of M31 and the Milky Way to be less than their known baryonic masses, or MOND requires all the galaxies in the Local Group to have non-radial orbits with respect to its center of mass.

In Section 2, I study the motion and timing between M31 and the Milky Way according to MOND and calculate their masses. In Sections 3 & 4, I study the motion and the timing of the dwarf galaxies in the Local Group according to Newtonian gravity and MOND respectively.

## 2. TESTING GRAVITY WITH MOTION BETWEEN M31 AND THE MILKY WAY

Kahn & Woltjer (1959) used an idealized model for the evolution of the Local Group. They modeled the Milky Way and M31 as two point masses that were initially moving apart with the Hubble flow but are now approaching due to their mutual gravitational attraction. With this radial orbit assumption, they estimated that the present separation and speed of the two galaxies required a total mass that was six times greater than the then known masses of these two galaxies. Binney & Tremaine (2008) repeated the calculation using modern values of the current separation (740 kpc), approaching speed between M31 and the Milky Way (125 km s$^{-1}$), and the age of the universe (13.7 Gyr), finding a total mass for the Local Group of about 4.6 × 10$^{12}$ M$_{sun}$. They also noted that the inclusion of dark energy increases the required mass by an additional small fraction. The conclusion is that Newtonian gravity requires a mass for the Local Group that is far larger than its baryonic mass of about 1.2 × 10$^{11}$ M$_{sun}$ (Binney & Tremaine 2008), which is often cited as a supporting argument for dark matter.

Here I rework this calculation assuming MOND. If $\mathbf{g_m}$ is the gravitational acceleration in MOND, and $\mathbf{g_n}$ is the Newtonian acceleration, Milgrom (1983) proposed that

$$\mathbf{g_m}\, \mu\, (\, |\, \mathbf{g_m}\, |\, /\, a_0\, )\, =\, \mathbf{g_n}\ , \qquad (1)$$

where $\mu(x) = 1$, when $x \gg 1$ (Newtonian regime),

$\mu(x) = x$, when $x \ll 1$ (Deep-MOND regime),

where $a_0$ is a new fundamental constant with the units of acceleration. Its value is approx 1.2 × $10^{-8}$ cm $s^{-2}$ (Sanders & McGaugh 2002).

In order to avoid elementary inconsistencies (e.g. Felten 1984), Bekenstein & Milgrom (1984) presented a rigorous Lagrangian formulation of MOND in which Poisson's equation is modified to become

$$\nabla \cdot [\, \mu \, (\, |\nabla \phi| \, / \, a_0) \, \nabla \phi \,] = 4\pi G \rho, \qquad (2)$$

where $|\nabla \phi| = g_m$.

This modification conserves linear momentum and energy as it is derived from Lagrangian formulation.

To study the motion of the M31 and the Milky Way system, the modified Poisson's Equation (2) should be used instead of the original MOND Equation (1). The original equation does not preserve the linear momentum for a two-body system except for the case where the two members of the two-body system have equal masses. However, there is no known analytical solution to the modified Poisson's equation. The equation needs to be solved numerically. To date no realistic simulations of two-body system in MOND exists (Zhao et al. 2010, hereafter Zhao10). Recently, Zhao10 derived an analytical expression for the two-body force in MOND-like theories from the viral theorem instead of solving the modified Poisson's equation. The two-body relative acceleration is given by (Zhao10, Equation 17)

$$\frac{d^2 r}{dt^2} = -\frac{G(m_1+m_2)}{r^2} - \frac{\sqrt{a_0 G}}{r} \frac{2(m_1+m_2)}{3(m_1 m_2)} \left[ (m_1+m_2)^{\frac{3}{2}} - m_1^{3/2} - m_2^{3/2} \right]$$

The acceleration is Newtonian ($\propto 1/r^2$) at short distances, and becomes MONDian ($\propto 1/r$) at large distances. Note that the above equation does not include the effect of the dark energy that is not significant in the M31 and Milky Way system (see Section 3 for detailed discussions).

The equation can be used to calculate the timing of the M31 and Milky Way system. The calculation has two parts. The first part covers the Newtonian regime and the second part covers the deep-MOND regime. With the realistic baryonic masses of M31 and the Milky Way, it can be shown that the time that the system is in the Newtonian regime is about two order of magnitude smaller than the time that the system is in the deep-MOND regime. Hence, we can use the approximate equation

$$\frac{d^2 r}{dt^2} = -\frac{\sqrt{a_0 G}}{r} \frac{2(m_1+m_2)}{3(m_1 m_2)} \left[ (m_1+m_2)^{\frac{3}{2}} - m_1^{3/2} - m_2^{3/2} \right]$$

to study the timing of the system.

This equation can be written as

$$\frac{d^2r}{dt^2} = -\frac{\sqrt{a_0 G m_1}}{r}\xi, \quad \xi \equiv \frac{2(m_1+m_2)}{3m_1^{3/2} m_2}\left[(m_1+m_2)^{\frac{3}{2}} - m_1^{3/2} - m_2^{3/2}\right], \quad (4)$$

where $r$ is the separation between M31 and the Milky Way, and $m_1$, $m_2$ are the masses of the Milky Way and M31 respectively. Note that $\xi$ is a dimensionless factor which depends on the relative magnitude of $m_1$ and $m_2$. It is interesting to point out that the same equation with a different scaling factor $\xi$ can be obtained by using the original MOND Equation (1) (Shi 2009).

With the current separation (740 kpc, Vilardell et al. 2009) and velocity (125 km s$^{-1}$, Binney & Tremaine 2008) between M31 and the Milky Way, and the age of the Local Group, I find the total mass of M31 and the Milky Way to be $5.63 \times 10^{10}$ M$_{sun}$. Appendix A shows the detailed calculation. Estimating M31 to be 50 per cent more massive than the Milky Way, I find the mass of the Milky Way to be $2.25 \times 10^{10}$ M$_{sun}$, which is only 45 per cent of the estimated baryonic mass of the Milky Way (about $5 \times 10^{10}$ M$_{sun}$, Binney & Tremaine 2008).

Instead of calculating the total mass, Equation (4) can be used to calculate the time required for M31 and the Milky Way to reach their current separation and velocity assuming known masses for M31 and the Milky Way. Using values for the baryonic masses of the Milky Way ($5 \times 10^{10}$ M$_{sun}$) and M31 ($1.5 \times 5 \times 10^{10}$ M$_{sun}$) from Binney & Tremaine (2008), I find that MOND predicts these two galaxies to have arrived at their current separation after 9 Gyr, which is shorter than the believed age of the Local Group.

Major uncertainties in the above argument arise because we do not know the exact past history of the Local Group. For example, if one assumes that M31 and the Milky Way had already completed one orbit and are approaching the completion of a second orbit now, using the known baryonic mass of M31 and the Milky Way as stated in the last paragraph the time required in MOND is 21 Gyr, which is again inconsistent with the current age of the Local Group. I find that the total baryonic mass of M31 and the Milky Way needs to be increased to $4.77 \times 10^{11}$ M$_{sun}$ in order for MOND to create a timing that is close to the age of the Local Group with the assumption of having completed one orbit. There are uncertainties on the baryonic mass of M31 and the Milky Way. For example, Tamm et al. 2007 estimated that the baryonic mass of M31 has a large uncertainty from 1 to $1.9 \times 10^{11}$ M$_{sun}$. And there is an uncertainty on the intergalactic gas between M31 and the Milky Way. Thus by just considering the M31 and the Milky Way system we cannot rule out the MOND prediction that the system has already completed one orbit with the nearly radial orbit assumption. However, in Section 4, I study the motion and timing of other dwarf galaxies in the Local Group. The much larger total baryonic mass of M31 and the Milky Way needed for creating the second passage is inconsistent with the observational data of the motion and timing of other dwarf galaxies.

Alternatively, we can drop the assumption that M31 and the Milky Way are moving on an almost radial orbit. A substantial component of non-radial motion can increase the estimated age from 9 Gyr (as predicted by MOND with the radial orbit assumption) to the age of the Local Group. The required transverse speed is of the order of 100 km s$^{-1}$, which is of the same order of magnitude as the current radial speed between M31 and the Milky Way. The corresponding angular speed of M31 would be $10^{-5}$ arcsec per year, which is not yet detectable but will be in the measurable range of SIM (Space Interferometry Mission) in the future (http://planetquest.jpl.nasa.gov/SIM; Peebles et al. 2001; Loeb et al. 2005).

van der Marel and Guhathakurta 2008 studied the observed velocities of the satellites of M31 and obtained a statistical determination of the transverse velocity of M31. They showed that the galactocentric transverse velocity of M31 is about 42 Km/s . This is in the right order of magnitude as the required transverse speed stated in the last paragraph. Thus, by just considering the timing of M31 and the Milky Way system we cannot rule out MOND. The only conclusion we can draw here is that the known baryonic masses of M31 and the Milky Way are inconsistent with a nearly radial orbit in MOND, and substantial transverse motion is required if MOND is correct.

## 3. TESTING NEWTONIAN GRAVITY WITH MOTION OF LOCAL DWARF GALAXIES

To study the motion of the dwarf galaxies in the Local Group rigorously, one needs to treat the system as a many-body system. Peebles used the Numerical Action Method (NAM) to trace dwarf galaxy orbits back in time (Peebles 1989, Peebles 1995). He treated Local Group members as point particles and also included the effect of external field due to larger scale structures by the neighboring group members, Maffei and Sculptor systems (Peebles 1989). His results indicate that the dwarf galaxies in the Local Group move on nearly radial orbits (Peebles 1995) which are consistent with the results obtained from the conventional N-body numerical integration forward in time.

Figure 1 shows the distance versus radial velocity with respect to the center of mass of M31 and the Milky Way for local dwarf galaxies using data taken from Fraternali et al. 2009. The plot does not show the satellite galaxies of M31 and the Milky Way which are bound to these two galaxies respectively. The circles represent those dwarf galaxies that belong to either the Local Group or to bridging regions between the Local Group and other nearby groups. The squares with distances greater than 2 Mpc represent galaxies that are not members of the Local Group. If the masses of M31 and the Milky Way were much smaller, most dwarf galaxies would scatter around the dashed line, which represents the Hubble flow. However, the expansion of the Local Group has been retarded by the gravitational attraction from M31 and the Milky Way, so that most dwarf galaxies have radial velocities smaller than predicted from the Hubble flow, as shown in Figure 1. Distant dwarf galaxies that are probably not members of the Local Group (open squares) can have velocities in excess of the Hubble flow if they have been pulled away from the Local Group toward other mass concentrations.

In the following, I use a simplified approach to gain some insight of the dynamics of dwarf galaxies in the Local Group. First, I assign the mass of the Local Group to M31 and the Milky Way. The Local Group has two mass centers which will deflect a dwarf galaxy out of the radial orbit if the dwarf galaxy initially moves radially outward with respect to the center of mass. However, for dwarf galaxies that move outward along the line linking M31 and the Milky Way or for dwarf galaxies that move radially outward with respect to the center of mass and perpendicular to the line linking M31 and the Milky Way, they will continue to move on radial orbits with respect to the center of mass. For dwarf galaxies that have greater distances from the center of mass than the separation between M31 and the Milky Way, they will also maintain almost radial orbits. I will use these dwarf galaxies as test particles to study their current radial velocity-distance relation as a function of the Local Group mass

Treating the Local Group as a point mass located at the center of mass of M31 and the Milky Way, the Newton's equation of motion for a test particle is

$$\frac{d^2 r}{dt^2} = -G \frac{(m_1 + m_2)}{r^2} , \qquad (5)$$

where $r$ is the distance of the test particle from the center of mass, and $m_1, m_2$ are the masses of the Milky Way and M31 respectively.

Again I assume that M31, the Milky Way, and the test particle were close to each other when the Local Group was formed and the test particle had an initial radial velocity with respect to the center of mass. The solid curve shows the radial velocity-distance relation of test particles at the present time, assuming a total mass for M31 and the Milky Way of $3 \times 10^{12}$ $M_{sun}$, and an orbit time of 13 Gyrs. The solid curve follows the approximate lower boundary of the dwarf galaxies in Figure 1. The dwarf galaxies close to the solid curve have nearly radial orbits with respect to the center of mass of the Local Group. Those above this curve have some transverse motion in their orbits.

The assumed total mass of $3 \times 10^{12}$ $M_{sun}$ for M31 and the Milky Way is perhaps 25 times their estimated baryonic mass (some $1.2 \times 10^{11}$ $M_{sun}$, Binney & Tremaine 2008). Thus here again we find that Newtonian gravity provides strong evidence for the existence of dark matter in the Local Group.

The above simplified analysis does not consider the effect of dark energy. With the inclusion of dark energy Equation (5) becomes

$$\frac{d^2 r}{dt^2} = -G \frac{(m_1 + m_2)}{r^2} + H_o^2 \, \Omega_\Lambda \, r ,$$

where $H_o$ is the present value of Hubble constant, and $\Omega_\Lambda$ is the density parameter for the cosmological constant (Peebles 2009).

With $H_o = 70$ Km s$^{-1}$ Mpc$^{-1}$, $\Omega_\Lambda = 0.7$ and the same assumptions used above, the corresponding radial velocity versus distance with respect to the center of mass is shown by the dotted curve in Figure 2. The inclusion of dark energy changes the motion slightly for dwarf galaxies which are close to the center of mass of M31 and the Milky Way. At 1 Mpc the effect of dark energy increases the timing calculation by 7%. Or equivalently, the total mass of M31 and the Milky Way needs to increase by 9% in order to create the same observed velocity at 1 Mpc at the present time. The effect of dark energy in the inner region of the Local Group is not significant which is consistent with the finding of Binney & Tremaine 2008.

## 4. TESTING MOND WITH MOTION OF LOCAL DWARF GALAXIES

In the deep-MOND regime, for high symmetry cases where the density distribution is spherical, or cylindrical, the modified Poisson's equation (Equation 2) reduces to Equation (1), which leads to a much simpler algebraic relation between the MOND acceleration and the Newtonian acceleration.

Thus, in the limit of low accelerations (deep-MOND regime), from Equation (1) the MOND acceleration is given by

$$g_m = \sqrt{a_0 g_n} \qquad (6)$$

This simplified equation is a good approximation for studying the motion of a test particle with respect to the center of mass of M31 and the Milky Way, since the test particle is in the deep-MOND regime most of the time except when the test particle is very close to M31 or the Milky Way. Note that the same simplified Equation (6) can be obtained from the relative acceleration (Equation 4) of a two-body system by imposing $m_1 \gg m_2$.

With the same assumptions as stated in Section 3, I repeat the calculation in MOND by using the simplified Equation (6) with the Newtonian acceleration $g_n$ given by Equation (5). The equation of motion of a test particle is

$$\frac{d^2 r}{dt^2} = -\frac{\sqrt{a_0 G (m_1 + m_2)}}{r},$$

where $m_1$, $m_2$ are the masses of Milky Way and M31 respectively, and $r$ is the distance of the test particle from the center of mass of M31 and the Milky Way.

Again I assume the orbit of the test particle is almost radial with respect to the center of mass of M31 and the Milky Way, the age of the Local Group is 13 Gyrs, and I use a mass of $1.2 \times 10^{11}$ M$_{sun}$ which is the estimated baryonic mass of M31 and the Milky Way. The dotted curve in Figure 1 shows the corresponding radial velocity-distance relation for MOND. The mathematics of the MOND calculations is similar to the calculation in Appendix A.

Note that the solid and dotted curves in Figure 1 are calculated with the simplifying assumption of a single point mass at the center of mass of M31 and the Milky Way, which breaks down when the distance to the test particle is comparable to the separation of M31 and the Milky Way. A slightly improved treatment is to regard the two galaxies as fixed separate masses, although in reality their separation changes over time. The inclusion of this small refinement leads to increased approaching speed at the same distance from the center of mass if the test particle is moving almost along the axis of M31 and the Milky Way, whereas a slight decrease in speed results if the test particle is moving at right angle to the axis.

For example, the dotted curve for MOND shows that the approaching speed of a test particle is 255 km s$^{-1}$ at a distance 600 kpc from the center of mass assuming M31 and the Milky Way are treated as a single point mass. If we treat M31 and the Milky Way as two separate masses with a fixed separation of 740 kpc and the test particle is moving almost along the axis of M31 and the Milky Way, the approaching speed obtained by MOND increases by 24 per cent which widens the separation between the MOND predictions and the data points of the dwarf galaxies in Figure 1. If the test particle is moving at the right angle to the axis, the approaching speed decreases by 6 per cent. This is a small change that the MOND predictions and the data points of the dwarf galaxies remain widely separated. Thus the simplifying assumption of a single point mass is justified.

The dotted curve shows that MOND predicts that all galaxies within 2 Mpc should be rapidly approaching the center of mass of M31 and the Milky Way at the present time if their orbits were purely radial. This is inconsistent with the observational data; there is no observed dwarf galaxy close to the dotted curve. Thus MOND requires that not a single dwarf galaxy in the Local Group has a nearly radial orbit with respect to the center of mass of M31 and the Milky Way.

I would have to reduce the baryonic mass of the Local Group by about an order of magnitude to $2 \times 10^{10}$ M$_{sun}$ in order to shift the dotted curve up to approximately the position of the solid curve in Figure 1. Thus radial orbits for local dwarf galaxies are inconsistent with MOND. This conclusion can be avoided with the following possibilities.

The first possibility: The distances to the dwarf galaxies are uncertain observational estimates, but would have to be underestimated by a factor of two or so to bring their locations in Figure 1 closer to the MOND curve. It seems unlikely that the measurements are all too low by such a large factor.

The second possibility: In Section 2 I discuss the possibility of the 2$^{nd}$ approach of the orbit of M31 and the Milky Way. It might be possible that the dwarf galaxies that are close to the center of mass of the Local Group have already completed one orbit with respect to the center of mass. For example, using the same much larger baryonic mass of $4.77 \times 10^{11}$ M$_{sun}$ which is needed to create a second passage of the Milky Way and M31 system as in Section 2, we could bring the MOND predictions of the current radial velocity-distance closer to the Solid curve in Figure 1 for the dwarf galaxies that have already completed one full orbit. However, in this case we should expect to see multiple turnaround radii (defined as the distances from the baryonic center of the Local Group at which the galaxies have a zero radial velocity) in Figure 1. Or, we should expect to see the velocities in Figure 1 change from approaching to receding then back to approaching to receding again as the distances of the dwarf galaxies increase. This is not consistent with the observational data. We see the velocities of the dwarf galaxies change monotonically from approaching to receding as the distances increase. Thus, from the observational data we can rule out the possibility that the dwarf galaxies in Figure 1 have completed one full orbit. The baryonic mass in the Local Group is not large enough for MOND to create a complete full orbit for dwarf galaxies near the center of mass. The same conclusion holds for the M31 and the Milky Way system.

The third possibility to save MOND is to drop the radial orbit assumption, and allow all the dwarf galaxies in the local universe to have substantial transverse motion. However, it is hard to explain how *all* the dwarf galaxies could have acquired considerable non-radial motion in our current view of the universe. When the Local Group was formed, the initial Hubble expansion will have given the dwarfs radially outward velocities; peculiar velocities, therefore, can have arisen only through gravitational interactions between galaxies. Since the largest two members of the Local Group are M31and the Milky Way, they pull the dwarf galaxies toward their center of mass, causing the largest component of the peculiar velocity to be in the radial direction. Interactions between neighboring dwarf galaxies must generate some transverse motion, which should be small due to the small relative masses of the dwarf galaxies except in rare cases where the neighboring dwarf galaxies are very close together. In other words, with the mass distribution dominated by M31 and the Milky Way it is unlikely that *all* the dwarf galaxies in the Local Group will have acquired transverse velocities comparable to their radial velocities.

We should also consider the effect of external field due to larger scale structures (LSS). For example, the Virgo and Coma clusters are about 60 Mpc and 100 Mpc respectively from the Milky Way. Famaey et al. 2007 (hereafter F07) studied the gravitational fields produced by these clusters as well as the Great Attractor. They found out that M31, the Virgo and Coma clusters, and the Great Attractor, all provide comparable gravitational fields at the Milky Way. The combined gravitational field produced by the LSS causes the Local Group to move with a 600 Kms$^{-1}$ flow with respect to the reference frame of the Cosmic Microwave Background (CMB). The tidal force from these structures will create relative motion of a dwarf galaxy with respect to the center of mass of the Local Group.

It is not an easy task to calculate the effect of the MOND tidal force from these distant sources. The total MOND force is not a linear sum of all contributors. We need to know the precise positions and masses of all the contributors to solve the nonlinear modified Poisson's equation. As stated in Section 2, there is no known analytical solution to the modified Poisson's equation. The equation needs to be solved numerically which is a complex task. In order to get some physical insight into the transverse motion of a dwarf galaxy due to the effect of the LSS, I adopt the simplified argument provided by F07. The combined MOND field from the LSS sources causes the Local Group to move with a speed of 600 Kms$^{-1}$ with respect to the CMB during a Hubble time. The order of magnitude of the acceleration is about $a_o/100$ (F07 Equation 9) which is about the same order of magnitude as the acceleration exerted by M31 on the Milky Way (F07 Equation 8). The MOND force from the LSS also creates relative motions between Local Group members. The transverse velocity of a dwarf galaxy with respective to the center of mass of the Local Group is determined by the strength of the gravitational field produced by the LSS at the Local Group multiplied by $\frac{\Delta d}{d} \times \sin(2\theta)$, where $\Delta d$ is the distance between the dwarf galaxy and the center of mass of the Local Group, $d$ is the distance between the LSS and the Local Group, and $\theta$ is the angle between the direction of the external acceleration and the line connecting the dwarf galaxy and the center of mass. Hence the effect of the tidal force in the transverse direction is about two orders smaller than the effect of the radial attraction from the Local Group assuming $\Delta d \sim 1$ Mpc and $d \sim 100$ Mpc. Thus, it is unlikely that **all** the dwarf galaxies in the Local Group will have acquired transverse velocities comparable to their radial peculiar velocities due to the effect of external field from the LSS even though that the gravitational field produced by the LSS is comparable to that of M31 or the Milky Way.

Furthermore, requiring all dwarf galaxies to have low anisotropy orbits is inconsistent with other MOND predictions. Klypin & Prada (2009, hereafter KP09) studied the motion of satellite galaxies around host galaxies using the observational data from the Sloan Digital Sky Survey (SDSS) (Adelman-McCarthy et al. 2006). They argued that the observed line of sight velocities are consistent with the standard cosmological model which predicts that the RMS velocities decrease with distance, and are inconsistent with MOND which predicts nearly constant line of sight velocity dispersions at larger distances. Angus et al. (2008, hereafter A08) disputed their conclusion, claiming instead that MOND can match the observed dispersion profile with a radially varying anisotropy $\beta(r)$. (The anisotropy parameter $\beta(r) = 1 - \sigma_t^2(r) / 2\sigma_r^2(r)$, where $\sigma_t$ is the transverse dispersion and $\sigma_r$ is the radial dispersion.) In order to match the observed decreasing line of sight velocity dispersions with distance, $\beta(r)$ needs to increase rapidly outwards and reach a value of 0.6 at 200 kpc and 0.8 at 500 kpc. (see Figure 1 of A08). Thus MOND requires high radial anisotropy orbits for satellites beyond 500 kpc. Since the dwarf galaxies in Figure 1 have a minimum distance of 500 kpc from the center of mass of M31 and the Milky Way, we expect A08 to predict $\beta(r) \sim 1$ for the galaxies at larger distances.

A08 also cited that their conclusion is consistent with the findings of Nipoti et al. (2007), who performed N-body simulations of elliptical galaxy formation in MOND. Their Figure 2 shows $\beta(r)$ increasing with distance from the center, and radial orbits predominate for stars at radii considerably greater than the half mass radius in the resulting elliptical galaxy. Thus, MOND predicts high radial anisotropy orbits for stars in elliptical galaxies as well as high radial anisotropy for satellite galaxies around host galaxies in the larger scale.

The high radial anisotropy at large distances in these two cases is inconsistent with my finding from the timing argument that the local dwarf galaxies require a large tangential bias in MOND. In short, either MOND predicts too little baryonic mass for M31 and the Milky Way, or MOND contradicts itself by demanding inconsistent radial anisotropy for different systems.

Note that the MOND timing calculations in Sections 2 and 4 do not include the effect of the external field due to the larger scale structures. Appendix B discusses the possible effect of the external field which does not change the conclusion in this Section.

## 5. SUMMARY AND CONCLUSION

I have reworked in MOND the original timing argument of Kahn & Woltjer (1959) for M31 and the Milky Way, and have found that either MOND predicts a total mass for M31 and the Milky Way that is smaller than the known baryonic mass in these two galaxies, or MOND requires a substantially non-radial orbit for this binary pair.

I have also tested MOND using the observed positions and velocities of dwarf galaxies in the local universe. Again I find that the predicted total mass for M31 and the Milky Way in MOND is much less than the known baryonic contents of these two galaxies with the nearly radial orbits assumption. Alternatively, using the known baryonic mass of M31 and the Milky Way, MOND does not allow any local dwarf galaxy to have a radial or nearly radial orbit about the center of mass of the Local Group. This requirement is inconsistent with MOND's prediction of a strong radial bias in the orbits of SDSS satellite galaxies and the motion of stars in the outer parts of elliptical galaxies. In short, the motion and timing of local dwarf galaxies might present problems for MOND.

MOND is extremely successful in explaining the flat rotation curves of spiral galaxies and the Tully-Fisher relation (Sanders & McGaugh 2002) without the need for dark matter. However, when it is applied to galaxy clusters, MOND predicts masses that are double (Sanders 2003) or even several times greater than the baryonic masses (Pointecouteau & Silk 2005). Sanders (2003) argued that the requirement of more mass for galaxy cluster does not constitute the falsification of MOND because there might be other forms of invisible matter (such as neutrinos) in the system. He argues that a definite falsification of MOND would arise when it predicts less mass than is observed, *i.e.* more mass can always be found but it is difficult to argue away observed mass. The timing analysis of the Local Group galaxies presented here come to the conclusion that either MOND predicts a total mass for M31 and the Milky Way that is less than the known baryonic contents of these two galaxies, or MOND requires that all the galaxies in the Local Group have non-radial orbits with respect to the center of mass of M31 and the Milky Way. The latter is inconsistent with other MOND's predictions of high anisotropy radial orbits in explaining the velocity dispersions of SDSS satellite galaxies and high anisotropy radial orbits for stars in elliptical galaxies at outer radii. The latter possibility can be tested with accurate measurements of the velocities and positions of local galaxies in the future, which will provide the best test for MOND against Newtonian gravity.

# APPENDIX A

# MOND TIMING CALCULATION

The MOND equation for the separation between M31 and the Milky Way is given by Equation (4) in the main text.

$$\frac{d^2 r}{dt^2} = - \frac{\sqrt{a_0 G m_1}}{r} \xi,$$

Where $\xi \equiv \frac{2(m_1+m_2)}{3 m_1^{3/2} m_2} [(m_1+m_2)^{\frac{3}{2}} - m_1^{3/2} - m_2^{3/2}]$. (A1)

At $r_m$ (the maximum separation between M31 and the Milky Way), $\dot{r} = 0$.

$$\dot{r}^2 = 2\xi \sqrt{a_0 G m_1} (\ln r_m - \ln r), \quad \text{or} \tag{A2}$$

$$\int \frac{dr}{\sqrt{\ln r_m - \ln r}} = -\sqrt{2\xi} (G m_1 a_0)^{1/4} \int dt \tag{A3}$$

The above equation can be integrated to obtain an explicit relation between r and t.

The total time for M31 and the Milky Way to move from $r = 0$ to $r = r_m$ and from $r = r_m$ to the current separation $r = r_n$

$t_{total} = t_{0m} + t_{mn} = 13$ Gyr (age of the Local Group).

In order to find the time $t_{0m}$ for M31 and the Milky Way to move from $r = 0$ to $r = r_m$, we need to perform the timing calculation in two parts. The first part covers the Newtonian regime from $r = 0$ to $r = r_t$. The second part covers the deep-MOND regime from $r = r_t$ to $r = r_m$. The transition separation, $r_t$ is the distance between M31 and the Milky Way at which the acceleration transits from the Newtonian regime to the deep-MOND regime. It can be shown that the time for M31 and the Milky Way to move from $r = 0$ to $r = r_t$ is about two orders of magnitude smaller than that required to move from $r = r_t$ to $r = r_m$ assuming the total mass is about the same order of magnitude as the known baryonic mass of M31 and the Milky Way.

Hence the total time $t_{0m}$ for moving from $r = 0$ to $r = r_m$, can be approximated by just integrating Equation (A3) from $r = 0$ to $r = r_m$.

$$t_{0m} = \frac{r_m \sqrt{\pi}}{\sqrt{2\xi}(G m_1 a_0)^{1/4}}.$$

The time for M31 and the Milky Way to move from $r = r_m$ to the current separation $r = r_n$

$$t_{mn} = \frac{r_m \sqrt{\pi}}{\sqrt{2\xi}(G m_1 a_0)^{1/4}} \; \text{erf}\left(\sqrt{\ln(\frac{r_m}{r_n})}\right).$$

The total time for M31 and the Milky Way to move from $r = 0$ to $r = r_m$ and from $r = r_m$ to the current separation $r = r_n$

$t_{total} = t_{0m} + t_{mn} = 13$ Gyr (age of the Local Group)

$$= \frac{r_m \sqrt{\pi}}{\sqrt{2\xi}(Gm_1 a_0)^{1/4}} \left[1 + \text{erf}\left(\sqrt{\ln(\frac{r_m}{r_n})}\right)\right] . \qquad (A4)$$

The maximum separation $r_m$ can be obtained by inserting the current separation $r_n$ (740 kpc) and the current speed $\dot{r}_n$ (125 km s$^{-1}$) between M31 and the Milky Way (Vilardell, Ribas & Jordi 2009, Binney & Tremaine 2008) into Equation (A2).

With $r_n$, $r_m$ and Equation (A4), we can find the MOND mass $m_1$ of the Milky Way numerically.

In order to find $m_1$ we need to provide a value for $\xi$. Estimating the mass of M31 to be about 1.5 times that of the Milky Way, we find $r_m = 1.03 \times 10^3$ kpc, and conclude the mass of the Milky Way $m_1 = 2.25 \times 10^{10}$ M$_{sun}$.

## Appendix B

## EXTERNAL FIELD EFFECT

As stated in Section 2, MOND is phenomenological rather than based on any stringent physical principles. In order to explain that there is no mass discrepancy for open clusters in the Milky Way, Milgrom 1983 proposed an additional phenomenological requirement that if a subsystem is placed in an external acceleration field $\mathbf{g_e}$, Equation (1) becomes (Sanders & McGaugh 2002)

$$\mathbf{g_m}\, \mu\,(\,|\,\mathbf{g_e} + \mathbf{g_m}\,|\,/\,a_0\,) = \mathbf{g_{n,sub}} \qquad (B1)$$

where $\mathbf{g_{n,sub}}$ is the Newtonian field of the subsystem alone. Thus, for a subsystem placed in a stronger external field where $|\mathbf{g_e}| \gg a_0$ the dynamics of the subsystem is Newtonian.

In an external field where $a_0 \gg |\mathbf{g_e}|$, the dynamics of the subsystem (Equation B1) behaves as follows as the radius increases from the center:

**I.** Newtonian regime: $\mathbf{g_m} = \mathbf{g_{n,sub}}$, when $|\mathbf{g_m}| \gg a_0 \gg |\mathbf{g_e}|$

**II.** MOND regime: $\mathbf{g_m}\,(\,|\mathbf{g_m}|\,/\,a_0) = \mathbf{g_{n,sub}}$, when $a_0 \gg |\mathbf{g_m}| \gg |\mathbf{g_e}|$

**III.** Anisotropic regime: $\mathbf{g_m} = (a_0\,/\,|\mathbf{g_e} + \mathbf{g_m}|)\,\mathbf{g_{n,sub}}$, when $a_0 \gg |\mathbf{g_m}| \approx |\mathbf{g_e}|$

**IV.** Enhanced G regime: $\mathbf{g_m} = (a_0\,/\,|\mathbf{g_e}|)\,\mathbf{g_{n,sub}}$, when $a_0 \gg |\mathbf{g_e}| \gg |\mathbf{g_m}|$

In the anisotropic regime the external field will enhance or decrease the MOND field depending on the angle between the force vectors of the internal and external fields. In the Enhanced G regime, the dynamics of the subsystem is Newtonian with a gravitational constant enhanced by a larger factor $a_0\,/\,|\mathbf{g_e}|$.

The corresponding modified Poisson's Equation (2) for a subsystem placed in an external field is given by (Famaey et al. 2007)

$$\nabla \cdot [\,\mu\,(\,|-\nabla\phi + \mathbf{g_e}|\,/\,a_0)\,\nabla\phi\,] = 4\pi G\rho \qquad (B2)$$

The dynamics of a dwarf galaxy in the Local Group can be obtained by solving the above equation with M31 and the Milky Way as the mass sources on the right hand side. This is a virtually impossible task as stated in Section 2.

Recently, Wu et al. 2007 (hereafter W07) studied the MOND dynamics of a Milky Way-like galaxy embedded in different external fields. They modeled the galaxy with a realistic representation of the Milky Way which has a thin disk, a thick disk, a spheroid, and a bulge similar to that of the Milky Way. With the model they solved Equation (B2) numerically. They found that the dynamics of the system transits from MONDian to Keplerian as the radius increases. For an external field of 0.01 $a_0$ the transition radius is ~ 500 kpc, while for an external field of 0.03 $a_0$ the transition radius is ~ 150 kpc . In the Keplerian regime, the potential at a large radius is nearly Newtonian with an enhanced factor $a_0 / | \mathbf{g_e} |$ (W07 Equation 4). This is consistent with Equation (B1) in the enhanced G regime.

The Local Group has two dominant members that make the dynamics of a dwarf galaxy in the Local Group more complicated than that in the Milky Way-like system. However, according to W07, the dwarf galaxies at 1~ 2 Mpc in Figure 1 are in the enhanced G regime. The dynamics of a dwarf galaxy in this regime is Newtonian with an enhanced gravitational constant. The same approach shown in Appendix A can be used to find the current radial velocity-distance relation for these galaxies as a function of mass. The calculation has two parts. One covers the MOND regime and the other covers the enhanced G regime. The required mass can be found numerically. Since the time that the system is in the MOND regime is shorter than the time that the system is in the enhanced G regime, as an approximation we can find the order of magnitude of the required mass by assuming the system is always in the enhanced G regime. As discussed in Section 3, the required mass to generate the radial velocity-distance relation (the solid curve in Figure 1) without considering the external field is $3 \times 10^{12}$ $M_{sun}$ . The effect of the enhanced gravitational constant is to reduce the required mass to generate the same field. With an external field of ~ 0.01 $a_0$, the required mass is reduced by about two orders of magnitude to $3 \times 10^{10}$ $M_{sun}$. This is about an order of magnitude less than the known baryonic mass $1.2 \times 10^{11}$ $M_{sun}$ in the Local Group. A similar conclusion holds for the M31 and the Milky Way system.

In summary, the inclusion of the external field effect due to the larger scale structures does not change MOND's underestimation of the Local Group mass.

# REFERENCES

Adelman-McCarthy, J., Agueros, M., Allam, S., Anderson, K., Anderson, S.,James, A.,Bahcall, N., Baldry, I., et al. 2006, ApJS 162, 38

Angus, G., Shan, H., Zhao, H., & Famaey, B. 2007, ApJ, 654, L13

Angus, G., Famaey, B., Tiret, O., Combes, F., & Zhao, H. 2008, MNRAS, 383, L1

Bekenstein, J., & Milgrom, M. 1984 ApJ, 286, 7

Binney, J., &Tremaine, S. 2008, Galactic Dynamics, $2^{nd}$ ed., Princeton University Press, NJ

Clowe, D., Bradac, M., Gonzalez, A., et al. 2006, ApJ, 648, 109

Famaey, B., Bruneton, J., & Zhao, H. 2007, MNRAS, 377, L79

Felten, J. 1984, ApJ, 286, 3

Fraternali, F., Tostoy, E., Irwin, M., & Cole, A. 2009, arXiv:0903.4635v1

Funkhouser, S. 2005, MNRAS, 364, 237

Gerhard, O., & Spergel, D. 1992, ApJ, 397, 38

Kahn, F., & Woltjer, L. 1959, ApJ, 130, 705

Klypin, A., & Prada, F. 2009, ApJ, 690, 1488

Komatsu, E., Dunkley, J., Nolta, M., Bennett, C., Gold, B.,Hinshaw, G., Jarosik, N., Larson, D., et al. 2008, arXiv:0803.0547v2

Loeb, A., Reid, M., Brunthaler, A., & Falcke, H. 2005, ApJ, 633, 894

Milgrom, M. 1983, ApJ, 270, 365

Milgrom, M. 2008, New Astronomy Reviews, 51, 906

Nipoti, C., Londrillo, P., & Ciotti, L. 2007, ApJ, 660, 256

Peebles, P.J. E. 1989, ApJ, 344, L53

Peebles, P. J. E. 1995, ApJ, 449, 52

Peebles, P. J. E. 2009, arXiv:0907.5207

Pointecouteau, E., & Silk, J. 2005, MNRAS, 364, 654

Rubin, V., & Ford, W. 1970, ApJ, 159, 379

Sanders, R., & McGaugh, S. 2002, ARA&A, 40, 263

Sanders, R. 2003, MNRAS, 342, 901

Sanders, R. 2008, arXiv:0806.2585S

Shi, Y. 2009, arXiv:0903.0591

Slosar, A., Melchiorri, A., & Silk, J. 2005, Phys Rev D, 72, 101301

Springel, V., Frenk, C., &White, S. 2006, Nature, 440, 1137

van der Marel, R., & Guhathakurta, P. 2008, ApJ, 678, 187

Vilardell, F., Ribas, I., Jordi, C., Fitzpatrick, E., & Guinan, E. 2009, arXiv:0911.3391

Wu, X., Zhao, H., Famaey, B., Gentile, G., Tiret, O., Combes, F., Angus, G., & Robin, A. 2007, ApJ, 665, L101

Zhao, H., Li, B., & Bienayme, O. 2010, arXiv:1007.1278

Zwicky, F. 1933, Helv. Phys. Acta, 6, 110

**FIGURES**

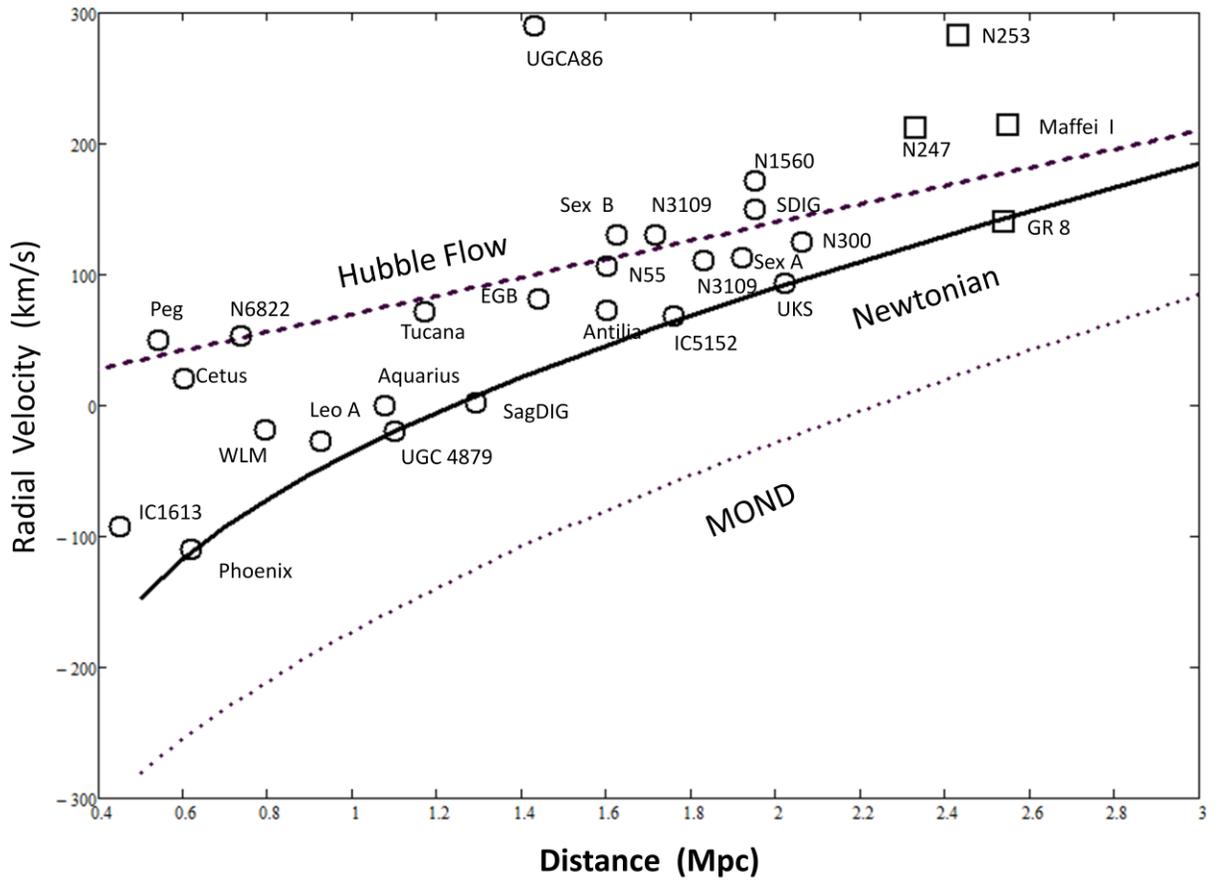

FIG. 1. The radial velocity versus distance with respect to the center of mass of M31 and the Milky Way for dwarf galaxies in the local universe. The plot does not show the satellite galaxies of M31 and the Milky Way which are bound to these two galaxies respectively.

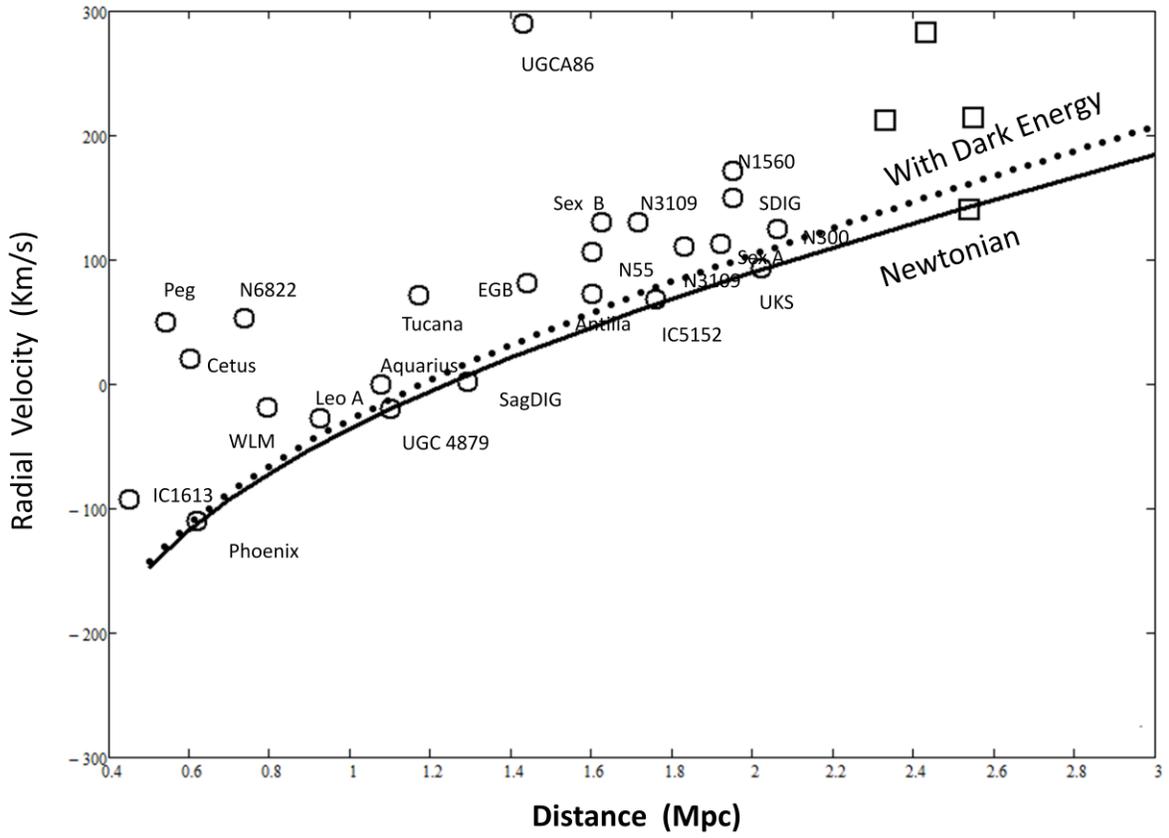

FIG. 2. Inclusion of dark energy changes the motion of dwarf galaxies in the Local Group slightly.